\documentclass[aps,prb,twocolumn,groupedaddress,superscriptaddress,showpacs]{revtex4}
\usepackage{graphicx}
\usepackage{color}

\begin{document}

\title{Adatoms in graphene as a source of current polarization: Role of the local magnetic moment}

\author{Matheus P. Lima}

\email[]{mplima@if.usp.br}
\affiliation{Instituto de F\'isica, Universidade de S\~ao Paulo, CP 66318, 05315-970, S\~ao Paulo, SP, Brazil.}

\author{Ant\^onio J. R. da Silva}
\email[]{ajrsilva@if.usp.br}
\affiliation{Instituto de F\'isica, Universidade de S\~ao Paulo, CP 66318, 05315-970, S\~ao Paulo, SP, Brazil.}
\affiliation{Laboratorio Nacional de Luz S\'incrotron - LNLS, CP 6192, 13083-970, Campinas, SP, Brazil.}

\author{A. Fazzio}
\email[]{fazzio@if.usp.br}
\affiliation{Instituto de F\'isica, Universidade de S\~ao Paulo, CP 66318, 05315-970, S\~ao Paulo, SP, Brazil.}

\date{\today}

\begin{abstract}

   We theoretically investigate spin-resolved currents flowing in large-area graphene, with and without defects,
doped with single atoms of noble metals ($Cu$, $Ag$ and $Au$) and $3d$-transition metals ($Mn$,$Fe$,$Co$ and $Ni$).
We show that the presence of a local magnetic moment is a necessary but not sufficient condition to have a non zero
current polarization. An essential requirement is the presence of spin-split localized levels near the Fermi energy
that strongly hybridize with the graphene $\pi$ bands. We also show that a gate potential can be used to tune the
energy of these localized levels, leading to an external way to control the degree of spin-polarized current without
the application of a magnetic field.

\end{abstract}


\maketitle

\section{Introduction}

Graphene\cite{graph-first,castro_RMP} possesses a very long mean free path of the order of micrometers 
and a ballistic transport regime at low temperatures\cite{5K}, which combined with a low spin-orbit
coupling make it ideal for applications in spintronic devices. However, due to the intrinsic 
spin-unpolarized electronic structure of a pristine sheet, it is necessary to modify the electronic
structure of graphene in such a way that the transport properties of the two spin channels become
different. Few ways that have been proposed to do this are (i) quantum confinement ({\it e. g.}
graphene nanoribbons\cite{nr1,nr2,nr3,nr4}, nanoflakes\cite{flakes}, hidrogenation\cite{singh}, and interaction
with substrates\cite{trench,sic-gap}); ii) strain\cite{strain}; and iii) adsorption of molecules or
atoms\cite{ads1,ads2}, where the presence of a local magnetic moment leads to a spin-non-degenerated
electronic structure. However, edge disorder is a limiting factor to use quantum confinement as a way to
differentiate the spin channels, and for the application of strain, there is the drawback of a mechanically
controllable device. Therefore, doping large-area graphene with magnetic dopants seems more suitable for
applications in spintronics.

In the last few years, motivated by spintronic applications, many studies focused on understanding the doping
properties of metallic adatoms on graphene. Experimentally, it has been shown that both noble metals and
transition metals tend to diffuse and clusterize in pristine graphene, unless they are trapped in lattice
defects such as monovacancies (MVs) and divacancies (DVs).\cite{exp1,exp2,exp3,exp4} Moreover, all theoretical
studies of metallic doping in large-area graphene have focused solely in investigating the emergence of local
magnetic moments\cite{th1,th1.1,th2,th3,th4,th5,th6,th7,th8,th9,th10,tm-vdW,th11,th12,th13}.
However, conclusions about the possibility of using metallic doping to generate spin-polarized transport requires
the investigation of spin-resolved currents, instead of merely local magnetic moments.

\begin{figure}[ht!]
\begin{center}
\includegraphics[width=8.5cm]{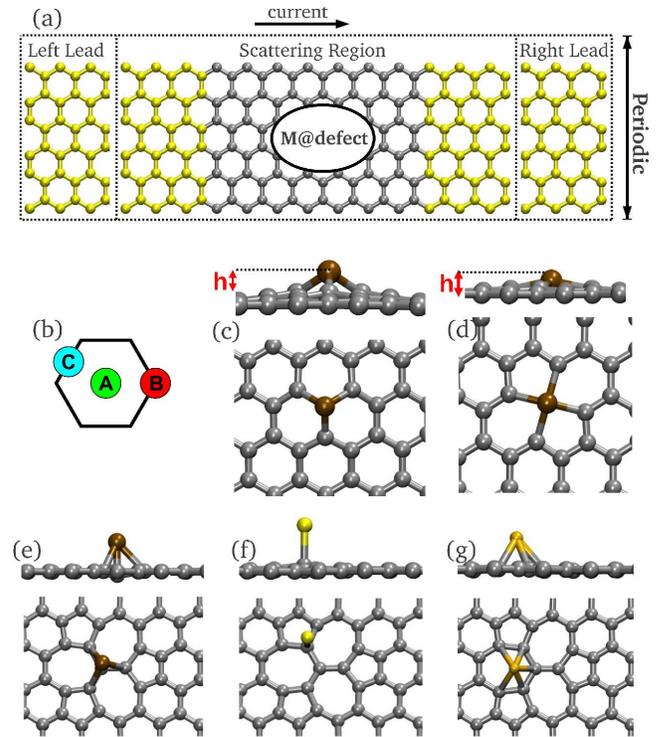}
\end{center}
\caption{\label{fig1} (Color Online)
(a) Representation of devices considered in this work.
(b) Possible adsorption sites for atoms on pristine graphene. (c) Geometry for $Co@MV$,
similar to all other atoms.  (d) $Co@585$, similar to all other atoms. (e) $Co@555777$ (similar to $Mn$:$Co$:$Ni@555777$).
(f) $Au@555777$. (g) $Cu@555777$ (similar to $Ag@555777$).}
\end{figure}

In this paper, we theoretically investigate spin-resolved currents flowing in large-area graphene, doped with single atoms of noble metals ($Cu$, $Ag$ and $Au$) and $3d$-transition metals ($Mn$,$Fe$,$Co$ and $Ni$). We study the
adsorption both in pristine graphene as well as in monovacancies and divacancies, and we show that the current polarization is not
monotonically correlated to the local magnetic moment. Instead, it depends on the presence of localized levels near the Fermi energy
which strongly couple (and thus hybridize) to the graphene $\pi$ bands. We also show that a gate potential can be used to tune the
energy of these localized levels, leading to an external way to control the degree of spin-polarized current without the application
of a magnetic field.

In Fig. \ref{fig1} (a) we show the setup of the devices we considered in this work, and in panels (b)-(g) we show the representative
geometries for the complexes of metals adsorbed in graphene, both pristine and with defects (metal@defect).
The paper is organized as follows: in the next section, we describe the methodology we used.
In the subsequent section, we present our results, discussing the total energy calculations, the current polarization,
the dependence on the doping concentration, and the effect of a gate potential. In the last section we present our conclusions.

\section{Methodology}

\begin{figure}[h!]
\begin{center}
\includegraphics[width=8.5cm]{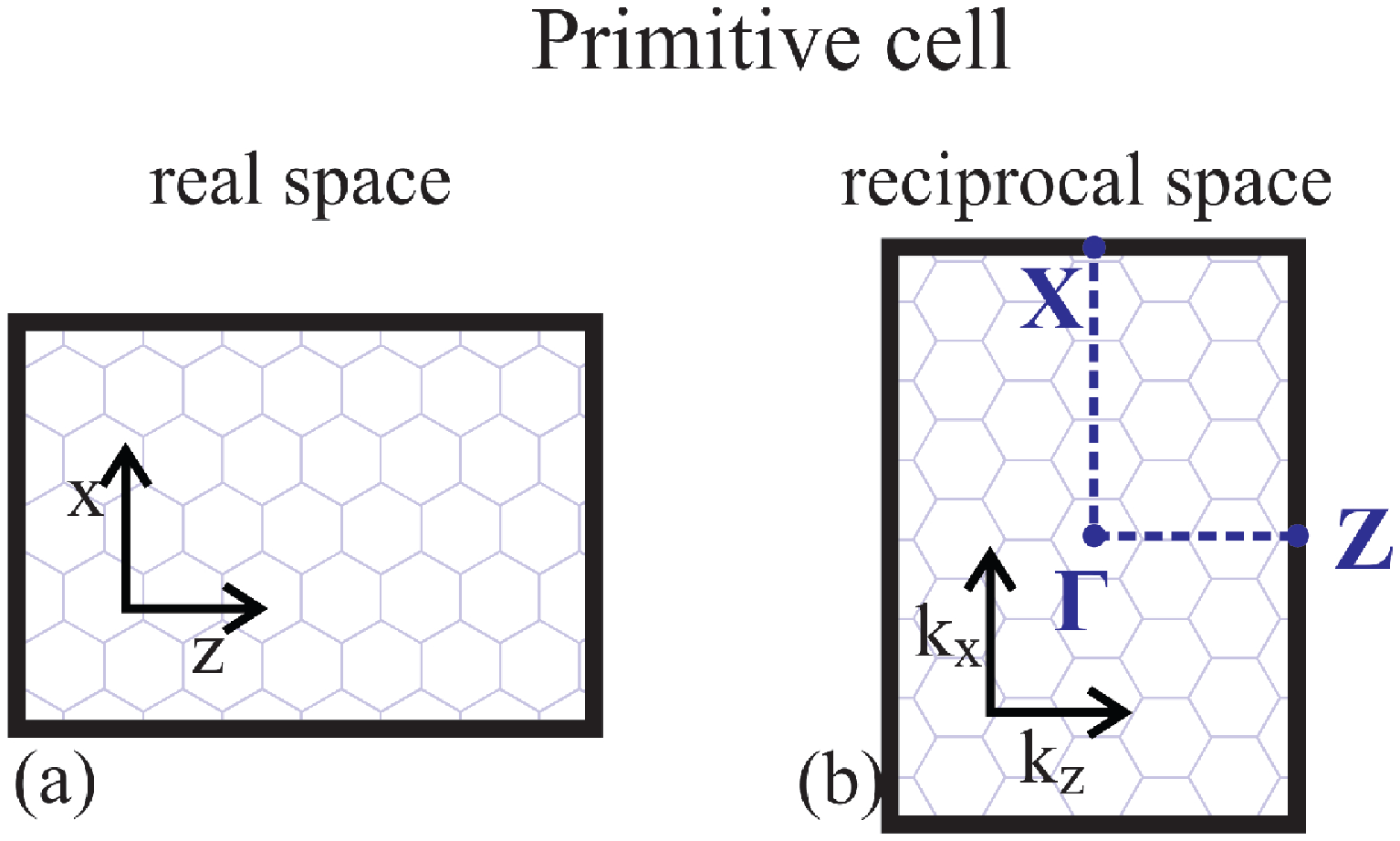}
\caption{\label{fig2} (Color Online) Schematic representation of the unitary cell in the real (a) and reciprocal (b) spaces, where we define
the symmetry points used to plot the energy bands.}
\end{center}
\end{figure}

In order to obtain the geometries and electronic structure, we employ total energy density functional theory\cite{dft} (DFT) calculations.
All geometries were fully relaxed using a force criteria of $0.02~$eV/\AA~
with the \textsc{Siesta} code\cite{siesta}.
All calculations were carried out applying the generalized
gradient approximation as parameterized by Perdew, Burke, and Ernzerhof\cite{gga-pbe} (GGA-PBE) for the exchange
correlation functional ($E_{xc}$). We do not consider the spin-orbit coupling, because such interaction may locally open small band
gaps\cite{so1,so2}, of the order of few milli electron volts, which will be a small effect when compared to the 
single-impurity scattering effects investigated in this paper.
A localized double-$\zeta$ polarized basis, an electronic temperature of $5meV$, and $300~Ry$ for the grid cut-off were used.
Most of our results were obtained with a rectangular $7\times6$ graphene supercell, as schematically shown in Fig. \ref{fig2}(a).
The integration in the Brillouin zone was done with the Monkhorst-Pack\cite{monkhorst} scheme corresponding to a $48\times48$ sampling
in the unitary cell. A vacuum of $20$\AA~ was used between the periodic images of graphene sheets
to avoid undesirable interactions. The directions between the symmetry points defined in Fig. \ref{fig2} (b) were used to plot the energy bands.

Once the relaxed geometries were obtained, we calculated the transport properties
with the \textsc{Transampa} code \cite{transampa}.
This code makes use of the non equilibrium Green's function techniques coupled to DFT
in a fully self-consistent way with the B\"uttiker-Landauer formalism\cite{negf1}
within the non interacting approach of Meir Wingreen\cite{negf4}.
Within this formalism, the translational symmetry along the transport direction is broken.
However, in the transversal direction, due to the periodicity of the infinite 2D-system, the Bloch theorem is still usable.
In order to take into account a truly 2D graphene sheet, we implement the transport equations
with transversal periodicity, considering $k$-points perpendicular to the transport direction ($k_\perp$).
In the transport calculations, the relaxed geometries were sandwiched between two buffer layers to guarantee that the scattering
region has a pristine-like coupling with the leads. We used
the same simulation parameters of the \textsc{Siesta} calculations, with exception of the $k$-points sampling.
Here, for the electronic density, we used the equivalent of 48 $k_\perp$-points in the unitary cell,
whereas for the transmittance, 5000 $k_\perp$-points are necessary for a smooth behavior. This number of
$k$-points is sufficient to capture the true 2D-character of graphene.
In each step of the self-consistent field cycle, we calculated the electronic density with an integral in
the complex plane discretized in 60 energy points and 5 poles. When a bias potential was applied,
the non equilibrium contribution for the electronic density was calculated with an integral
discretized in 200 energy points.

We also implement the effect of a gate potential with strength "$V_{gate}$" by adding to the Hamiltonian a function $V(z)$
that depends only on the $z$ coordinate, given by:
\begin{widetext}
\begin{eqnarray}
V(z)=\left\{\begin{array}{ll} 0 &, z<z_1 \text{ or } z>z_2 \\
                                  \frac{V_{gate}}{2}\left\{ 1-cos\left[ \frac{\pi(z-z_{1})}{\tau} \right] \right\} &, 0<z-z_{1}<\tau \\
                                  V_{gate} &, z_{1}+\tau<z<z_{2}-\tau \\
                                  \frac{V_{gate}}{2}\left\{ 1+cos\left[ \frac{\pi(z_2-\tau-z)}{\tau} \right] \right\} &, -\tau<z-z_2<0.\\ \end{array}\right.
\end{eqnarray}
\end{widetext}
Since $z$ is the transport direction, the above function affects only a region between $z_1$ and $z_2$ (gated region) and
is continuous up to its first derivative. This region encloses the metal and the defect in our simulations, and we verified
that the buffer-layer regions were not affected by this gate.
The $\tau$ parameter controls the
smoothness of the connections between the gated and non-gated regions, and
we used $\tau=1.25~$\AA~ for a smooth behavior.
It is also important to mention that the gate effect is included in the whole self-consistent-field procedure, and the electronic density
rearrangement creates a screening effect, in the sense that (for example) $V_{gate}=1V$ will not necessarily shift the energy levels by $1eV$.
This implementation is in the same spirit of Refs. \onlinecite{gate1} and \onlinecite{gate2}.

\section{Results and Discussion}

\subsection{Geometry, magnetic moment, and binding energies}

\begin{table*}
\caption{\label{tab1} Binding energies ($E_b$) in $eV$, magnetization $m$ in $\mu_B$, and metal-carbon bound lengths ($d_{MC}$) in \AA~ for
all geometries considered in this work. $d_{MC}$ are shown between brackets when there are different lengths.
When convenient, we show either the position (see Fig. 1) or the local symmetry group.
The binding energy is defined as $E_{b}^{atom@defect}=E_{total}^{atom@defect}-E_{total}^{atom}-E_{total}^{defect}$}
\begin{ruledtabular}
\begin{tabular}{ccccc}
       &   M@pristine   &   M@MV     &    M@585    &    M@555777  \\
\begin{tabular}{c}
\hline
atom \\
\hline
Mn   \\
Fe   \\
Co   \\
Ni   \\
Cu   \\
Ag   \\
Au   \\
\end{tabular}

&

\begin{tabular}{cccc}
\hline
$E_b$  & $d_{MC}$ &   $m$   &  pos.\\
\hline
-0.189 &   2.53   &   5.5   &  A \\
-0.874 &   2.10   &   2.0   &  A \\
-1.33  &   2.10   &   1.1   &  A \\
-1.83  &   2.11   &   0.0   &  A \\
-0.240 &   2.09   &   1.0   &  B \\
-0.090 &   3.34   &   1.0   &  B \\
-0.410 &   2.39   &   1.0   &  B \\
\end{tabular}

&

\begin{tabular}{cccc}
\hline
$E_b$  & $d_{MC}$ &   $m$   &  symm.  \\
\hline
-6.02  &   1.84   &   3.0   &   $C_{3v}$ \\
-7.19  &   1.77   &   0.0   &   $C_{3v}$ \\
-7.89  &   1.77   &   1.0   &   $C_{3v}$ \\
-7.26  &   1.80   &   0.0   &   $C_{3v}$ \\
-3.84  &   $(1.92/1.89/1.89)$   &   1.0   &   $C_{s}$ \\
-1.97  &   $(2.23/2.19/2.19)$   &   1.0   &   $C_{s}$ \\
-2.57  &   $(2.09/2.13/2.13)$   &   1.0   &   $C_{s}$ \\
\end{tabular}

&

\begin{tabular}{ccc}%c}
\hline
$E_b$  & $d_{MC}$ &   $m$  \\%&  symm.  \\
\hline
-5.14  &   1.97   &   3.1  \\%& $C_{2v}$   \\
-6.09  &   1.96   &   3.9  \\%& $C_{2v}$   \\
-6.49  &   1.92   &   1.1  \\%& $C_{2v}$   \\
-7.01  &   1.88   &   0.0  \\%& $C_{2v}$   \\
-5.03  &   1.92   &   0.0  \\%& $C_{2v}$   \\
-2.54  &   2.00   &   0.0  \\%& $C_{2v}$   \\
-4.30  &   2.00   &   0.0  \\%& $C_{2v}$   \\
\end{tabular}

&

\begin{tabular}{cccc}
\hline
$E_b$  & $d_{MC}$ &   $m$   &  pos.  \\
\hline
-1.28  &   2.25   &   5.8  & fig. \ref{fig1} (e) \\
-1.84  &   1.89   &   3.1  & fig. \ref{fig1} (e) \\
-2.26  &   1.82   &   1.7  & fig. \ref{fig1} (e) \\
-2.44  &   1.82   &   1.2  & fig. \ref{fig1} (e) \\
-1.55  & ${2.37/2.37/2.20 \choose 2.18/2.18}$ &   0.0  & fig. \ref{fig1} (g) \\
-0.810 & ${2.86/2.70/2.63 \choose 2.53/2.45}$ &   0.0  & fig. \ref{fig1} (g) \\
-1.15  &   2.12   &   0.0  & fig. \ref{fig1} (f) \\
\end{tabular}

\\
\end{tabular}

\end{ruledtabular}
\end{table*}

The relaxed geometries, local magnetic moments ($m$) and binding energies ($E_b$)
are shown in table \ref{tab1}. Our results compare well with previous
calculations\cite{ads2,th1.1,th2,th3,th4,th5,th6,tm-vdW}. This is also true for the
adsorption of noble metals on pristine graphene, where van der Waals interactions are
important\cite{tm-vdW}. However, in these cases, the energies obtained for the different
configurations with our methodology are all very similar. Thus, to select which adsorption
site to use in the charge transport calculations [B position in Fig. \ref{fig1}(b)], we
used the results of Ref. \onlinecite{tm-vdW}. For atoms trapped in the MV, all relaxations
lead to geometries similar to $Co@MV$ [depicted in Fig. \ref{fig1} (c)]. However, the
transition metals have a $C_{3v}$ local symmetry, whereas for the noble metals this symmetry
is destroyed by a structural distortion, breaking the spin degeneracies of localized states
near the Fermi level, leading to $m=1.0~\mu_B$, as discussed in Ref. \onlinecite{th2}.
For adsorption on DVs, we considered two particular reconstructions studied in the
literature\cite{digao}, the 585 and 555777. For atoms trapped in 585 DVs, we find a
$C_{2v}$ local symmetry in all cases, with the atoms displaced from the graphene plane,
similar to $Co@585$, as shown in Fig. \ref{fig1} (d). All the $3d$ transition metals adsorbed
on the 555777 DV present a non-zero local magnetic moment, with the energetically most
favorable position being above the central atom of this defect, as shown in Fig. \ref{fig1} (e).
It is worth noting that (i) $Ni$ has a non zero local magnetic moment solely when trapped in
this site and (ii) adsorption of adatoms has been experimentally observed in them\cite{exp3}.
The noble metals, on the other hand, have a completely different behavior, presenting a
zero local magnetic moment when adsorbed on the 555777 DV. The lowest energy configuration
for $Au@555777$ is shown in Fig. \ref{fig1} (f), with the gold atom at the top of a carbon
atom belonging to a pentagon. For $Cu$ and $Ag$, the lowest energy position is above the
center of a heptagon, as shown in Fig. \ref{fig1} (g). However, $Cu$ presents a $C_{s}$ local
symmetry, whereas $Ag$ presents five different chemical bond lengths.

\subsection{Current Polarization}

\begin{figure}
\begin{center}
\includegraphics[width=8.5cm]{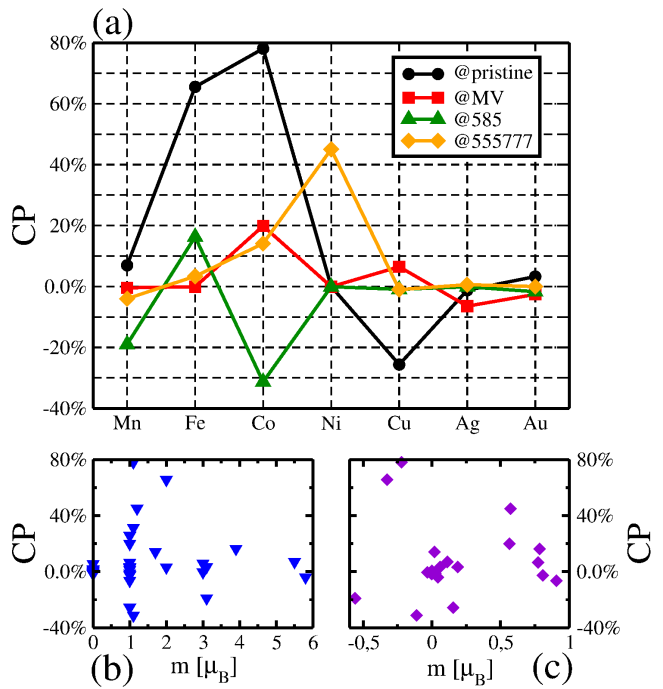}
\end{center}
\caption{\label{fig3} (Color online) Current polarization with $V_{bias}=50meV$ (a) for all studied cases,
(b) as a function of the local magnetic moment, (c) as a function of the magnetic moment induced
at the graphene carbon atoms. The solid lines guide the eye.}
\end{figure}

In Fig. \ref{fig3} (a) we show the current polarization, CP=$(I_\uparrow-I_\downarrow)/(I_\uparrow+I_\downarrow)$,
for all studied cases, where the current for each spin channel ($I_\sigma$) is
$I_\sigma= (e/h)\int  T_\sigma (E) [f(E-\mu_L)-f(E-\mu_R)]dE$, with $T_\sigma(E)$
the transmittance for the spin $\sigma$, $f(E)$ the Fermi-Dirac function,and $\mu_{L(R)}$
the chemical potential for the left (right) electrodes. A bias potential of $50~meV$ was used.
Clearly, $Co$ and $Fe$ adsorbed on pristine graphene generate the higher CP in the systems we consider.
However, the metal diffusion experimentally observed creates a drawback to build devices with
adatoms over pristine graphene since in the most cases the atoms tend to clusterize. For the atoms
trapped in defects, the higher values of CP (in modulus) occurs for the $Ni@555777$, followed by
$Co@585$. The last one has a negative CP because localized levels above the Fermi-energy scatter
majority spins, generating a current with excess of minority spin electrons. One should note that,
despite the high magnetization of all $Mn$ complexes, the absolute values of CP are not so remarkable in such systems.
In particular, the $Mn@MV$ has a negligible CP, even though it has $m=3.0~\mu_B$. For the noble metals,
an unusual local magnetic moment generated by the polarization of "$s$" atomic orbitals leads
to CP smaller than $10\%$. These results indicate that the existence of a local magnetic moment is a necessary
but not sufficient condition to have a large CP. This can be clearly seen in the Fig. \ref{fig3} (b),
where we show the CP as a function of the local magnetic moment,
showing that there is no correlation between the local magnetic moment and the current polarization.
In Fig. \ref{fig3} (c) we show the CP as a function of the magnetic moment induced only in the graphene carbon atoms.
This magnetization occurs due to the hybridization between the graphene states and the adatom orbitals. However,
this hybridization may occur (i) at deep levels, not affecting the CP, or (ii) near the Fermi energy,
affecting the transmission probability near the transport window. Therefore we conclude that the
CP behavior is not governed solely by the  magnetic moment induced in graphene, but depends fundamentally 
on the energy window where the hybridization responsible for the appearance of such a magnetic moment occurs.

\begin{figure}
\begin{center}
\includegraphics[width=8.5cm]{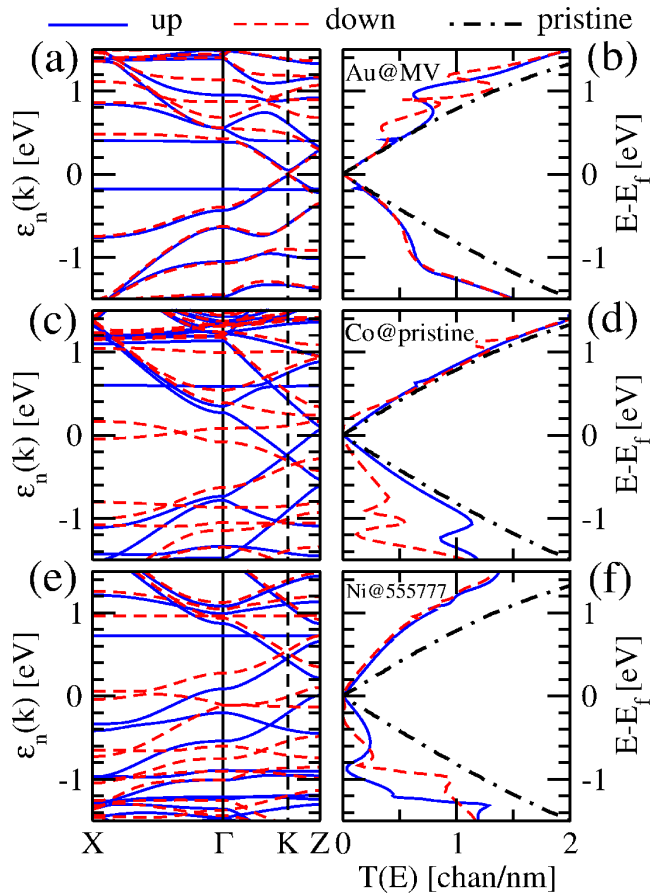}
\end{center}
\caption{\label{fig4} (Color online) Band structure and spin-resolved transmittance for
(a) and (b) $Au@MV$, (c) and (d) $Co$ over pristine graphene, (e) and (f) Ni@555777. The solid (dashed) lines
correspond to the majority (minority) spin channels. For the sake of comparison, we also show the
transmittance for pristine graphene. The smooth behavior of all curves indicates the convergence,
regarding the $k_\perp$-points, to the true 2D-nature of graphene. In the energy bands, we also indicate the 
position of the Dirac cone (with a vertical dashed line),
corresponding to the $K$ symmetry point in a hexagonal supercell.
}
\end{figure}

In order to understand the ruling factors that govern the current polarization,
we performed band structure calculations. From these results,
we conclude that the degree of current spin polarization is governed by (i) the presence of localized
levels, non degenerated in spin, nearby the Fermi level and (ii) by their hybridization with graphene levels.
To illustrate these conclusions we present in detail three representative cases shown in Fig. \ref{fig4}.
In panel (a) we plot the energy bands for the $Au@MV$ complex. In this situation, there is a flat band around
$0.2~eV$ below the Fermi energy, with majority spin associated with Au "s" atomic orbitals, which is
responsible for the local magnetic moment of $1.0~\mu_B$. This band does not modify the shape of the
Dirac cone because it has a weak coupling to the graphene $\pi$ bands. As a consequence, the transmission
probability [$T(E)$] shown in Fig. \ref{fig4} (b) is very similar to the pristine case around the Fermi energy.
This example shows that a magnetic moment and localized states near the Fermi energy are a necessary, but not a
sufficient, condition to generate spin-polarized currents in graphene, as discussed above. In fact, to generate
current polarizations it is also necessary to have a high coupling between the localized levels and graphene
$\pi$ bands, modifying in this way the transmission probability for only one or both spin channels. As an example
of modifications in only one spin channel, we show the energy bands for the $Co$ adsorbed on pristine graphene
in Fig. \ref{fig4} (c). It is notable that a Dirac cone is well defined only for the majority spin. For the
minority spin, localized levels hybridize with the Dirac cone states, and hence significantly modify the dispersion
of these states. As a result, the transmittance [shown in Fig. \ref{fig4} (d)] is very similar to the one in pristine
graphene only for the majority spin, causing in this way a high current polarization. For the $Ni@555777$ complex
there are modifications in the band structure for both spin channels [see Fig. \ref{fig4} (e)]. In this case, the
local magnetic moment comes from localized states around the Fermi energy with a high hybridization with the Dirac cone.
As a consequence, the transmittance for both spin channels, shown in Fig. \ref{fig4} (f), are not only distinct
from the pristine case, but also very different between them. In this system we still observe a flat band at
$0.75~eV$ above the Fermi energy. However, as there is a weak coupling with the Dirac cone states, there is almost
no change in $T(E)$. Therefore, the examples shown in Fig. \ref{fig4} elucidate the conditions to have 
spin-polarized currents in graphene generated by doping with magnetic impurities.

\subsection{Dependence on the doping concentration}

\begin{figure}[h!]
\begin{center}
\includegraphics[width=8.5cm]{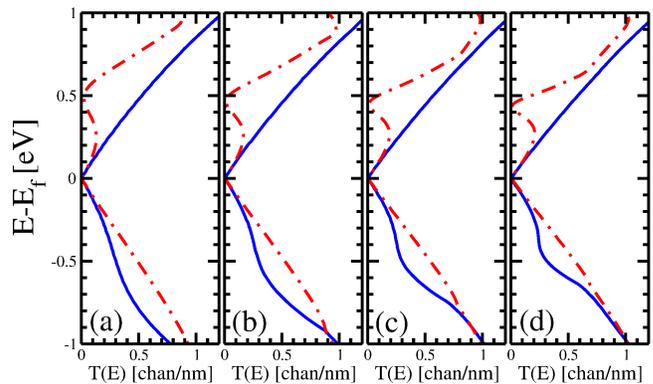}
\end{center}
\caption{\label{fig5} (Color online) Spin-resolved transmittance for $Mn@MV$. The distance
between $Mn$ atoms are $13.0$, $17.3$, $21.6$ and $26.0~$\AA~ in (a)-(d), respectively.}
\end{figure}

In most calculations we performed the distance between the metallic adatom and its periodic image is $13.0~$\AA. In order to understand
the effect of the doping concentration we calculated the transmittance for $Mn@MV$ also for supercells of $17.3$, $21.6$, and $26.0~$\AA~ wide,
as shown in Fig. \ref{fig5}. The main characteristic of $Mn@MV$ transmittance is a valley in the minority spin transmittance around $0.5~eV$
above the Fermi energy, and another valley for the majority spin transmittance around $0.7~eV$ below the Fermi energy. With the widening of the
supercell, the valleys became more localized, as a reflect of the decrease in the energy dispersion of the defect levels.
However, the main results obtained with the smaller supercell are unchanged.

\subsection{Effect of a gate potential}

\begin{figure}[ht!]
\begin{center}
\includegraphics[width=8.5cm]{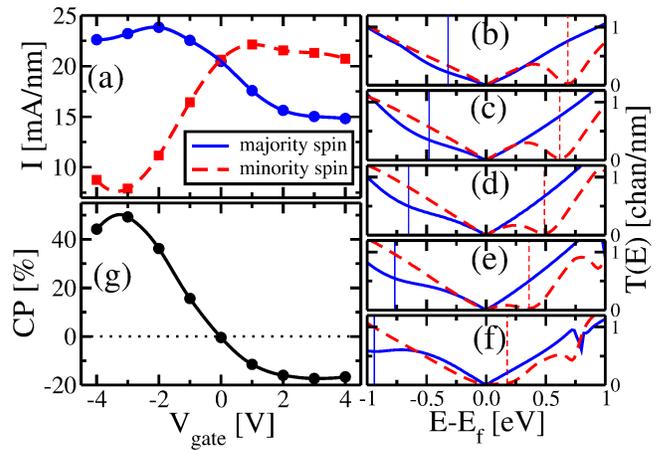}
\end{center}
\caption{\label{fig6} (Color online) (a) Spin-resolved current for the $Mn@MV$ complex,
were the circles (squares) represent the majority (minority) spin.
(b)-(f) Transmission probability for
$V_{gate}=3.0$, $1.0$, $0.0$, $-1.0$ and $-3.0~V$, respectively.
Vertical lines indicate the position of the valleys in the transmittance. The Fermi energy is fixed by the infinite graphene leads.
(g) CP as a function of $V_{gate}$.}
\end{figure}

We also show that it is possible to control the current polarization via the effect of a gate potential.
As an example, we show in Fig. \ref{fig6} the transport properties for the $Mn@MV$ complex, as a function of
$V_{gate}$. In panel (a) we present the spin-resolved currents. Without a gate potential the current polarization is
negligible. However, with the inclusion of a negative gate the current has excess of majority spin electrons,
whereas with a positive gate there are excess of minority spin electrons. This behavior occurs because the gate
potential shifts the localized states that hybridize with the graphene $\pi$ bands, as shown in Figs. \ref{fig6}(b)-\ref{fig6}(f),
allowing a tuning of the valleys in the transmittance (indicated by vertical lines). In Fig. \ref{fig6}(g), we also present
CP as a function of $V_{gate}$. It is remarkable that CP can be changed from $50\%$ to $-18\%$ with the application of a gate.
Therefore, without any external magnetic perturbation, it is possible to control the degree of spin polarization in such systems.

\section{Conclusion}

In conclusion, via studies of adsorbed metal atoms in graphene with and without defects, we show that the presence
of a local magnetic moment is a necessary but not sufficient condition to create a current polarization. Essentially,
a fundamental requirement to have a non-zero current polarization is the presence of spin-split localized levels near
the Fermi energy that strongly hybridize with the graphene Dirac-cone states.
We also show that with an external gate potential it is possible to control the degree of current polarization without
any magnetic perturbation. The knowledge of such properties for isolated scatterers is a fundamental first step to design
spintronic devices, where other important effects such as multiple scattering and disorder may even enhance the
polarization signal\cite{multiple1,multiple2,multiple3}.

The authors acknowledge the financial support from the Brazilian funding agencies INCT/CNPq and FAPESP.

%\bibliography{ref}

\end{document}